# Molecular Packing Analysis of the *Smectic E* Phase of a Benzothieno-benzothiophene Derivative by a Combined Experimental / Computational Approach

Sebastian Hofer[a], Wolfgang Bodlos[a], Jiří Novák[b], Alessandro Sanzone[c], Luca Beverina[c], Roland Resel[a]*

[a]*Institute of Solid State Physics, Graz University of Technology, Austria;* [b]*Department of Condensed Matter Physics, Masaryk University, Brno, Czech Republic,* [c]*Department of Materials Science, University of Milano-Bicocca, Milano, Italy*

* roland.resel@tugraz.at

# Molecular Packing Analysis of the *Crystal E* Phase of a Benzothieno-benzothiophene Derivative by a Combined Experimental / Computational Approach

## Abstract

The molecule 2-decyl-7-phenyl[1]benzothieno[3,2-b][1]benzothiophene has gained a lot of attention, since high charge carrier mobility was observed in thin film transistors. Its thermotropic liquid crystalline states may play an important role in the thin film formation, since the *smectic A* and the s*mectic E* phase (SmE) are claimed to be pre-stages of the final bulk structure. To understand the phase diversity, structural characterisation of solution processed thin films are performed by X-ray diffraction in the complete temperature range up to the isotropic state at 240°C. The diffraction pattern of the SmE phase is analysed in detail. Peak broadening analysis reveals that the crystallographic order across the smectic layers is larger than the order along the smectic layers. A combined experimental and computational approach is used to determine the molecular packing within the SmE phase. It leads to a number of different packing motifs. A comparison of the calculated diffraction pattern with the experimental results reveals that nano-segregation is present within the SmE phase. Energy consideration clearly favours a herringbone arrangement of the aromatic units. The nano-segregation within the SmE phase with herringbone packing of the aromatic units is accompanied with interdigidation of sidechains from neighbouring smectic layers.

**Introduction**

The class of alkyl substituted benzothienobenzothiophene type molecules gained considerable attention due to high charge carrier mobilities in organic thin film transistors. Values of up to 9 cm$^2$V$^{-1}$s$^{-1}$ are observed [1][2]. The presence of alkyl chains symmetrically attached at the terminal ends of the molecule provides overall flexibility in device fabrication, since thin films can be prepared by solution processing [3][4] as well as by physical vapour deposition [5][6]. Moreover, the thermotropic liquid crystalline state can be used in the thin film preparation procedure to obtain a defined structure within thin films [7][8].

Recently, a new asymmetric derivative of BTBT – the molecule 2-decyl-7-phenyl[1]benzothieno[3,2-b][1]benzothiophene (Ph-BTBT-10) - was introduced [9]. The asymmetry in the substitution pattern, featuring a phenyl residue on the 2 - position of the BTBT core and a decyl chain on the 7 – position, results in a more divers phase behaviour[10]. Liquid crystalline states with high structural order appear. It is reported that these states can be used as pre-stages of the final bulk crystallisation so that crystals with extended size can be prepared within thin films [11].

The asymmetry of the molecule causes crystallisation with a bilayer herringbone structure [9,10]. The structure is formed by stacked individual layers consisting of aromatic units arranged in a herringbone manner and decyl chains formed by the terminal alkyls. Two herringbone layers stack directly on top of each other leading to double herringbone layers. The crystal structure can be described by a sequence of these double herringbone layers followed by double layers of the decyl chains. The lattice constants of the structure are *a = 6.047 Å, b = 7.757 Å, c = 53.12 Å* and *β = 93.14°* whereby *c* spans over two herringbone layers and two alkyl layers [12]. In summary the molecules show a *head-to-head* arrangement (along the c-axis), since the aromatic units of molecules from neighbouring herringbone layers are antiparallel to each other. A phase transition to *smectic E* (SmE) with a *head-to-tail* arrangement of the molecules is observed at temperatures around 144°C [9][10]. The presence of a SmE phase was concluded by polarized optical microscopy investigations [13]. Analysis of the molecular packing at elevated temperatures is already performed by molecular dynamics simulations suggesting a structure built by stacked single molecular sheets in a *head-to-tail* arrangement. However, the arrangement of the molecules within the SmE phase is still not clear [14][15][16].

Recent developments in the field of crystal structure solution from thin films allow the determination of the molecular packing within unknown crystal structures exclusively present within thin films [17][18]. Grazing incidence X-ray diffraction in combination with specular X-ray diffraction are used for this purpose. Despite the limited information of only few diffraction peaks, a combination of experimental performance and theoretical calculations can be suitable to determine the molecular packing [19]. Here, we will apply our knowledge on the crystal structure solution from thin film to the SmE phase of the molecule Ph-BTBT-10.

**Experimental**

**Thin film preparation**

The molecule Ph-BTBT-10 was synthetized according to the strategy shown in Scheme 1 that some of us recently published [20]. The commercially available [1]benzothieno[3,2-b][1]benzothiophene (BTBT) core was regioselectively acylated at the 2 position via a Friedel Crafts reaction with decanoyl chloride. The obtained ketone was converted to 2-decyl-[1]benzothieno[3,2-b][1]benzothiophene (C10-BTBT) through reduction with a $NaBH_4/AlCl_3$ mixture in good yield. Bromination of the latter with bromine in chloroform afforded 2-decyl-7-bromo-[1]benzothieno[3,2-b][1]benzo-thiophene (Br-BTBT-10) after chromatographic purification. Finally, a Suzuki-Miyaura coupling between Br-BTBT-10 and phenyl boronic acid in a Kolliphore EL 2 wt%: toluene emulsion (9:1 v/v) in the presence of triethyl amine as the base and [1,1′-Bis(di-tert-butylphosphino)ferrocene]dichloropalladium(II) Pd(dtbpf)$Cl_2$ as the catalyst gave 10-BTBT-Ph in essentially quantitative yield.

Thin film samples were prepared via spin-coating or drop-casting from toluene solutions of varying concentrations and spin speeds (0.2 g/l up to 5.0 g/l and 500 rpm up to 3000 rpm) onto thermally oxidised silicon wafers. Powder samples were prepared by transferring a small amount of about 1mg of polycrystalline material onto a silicon substrate and a subsequent flattening procedure.

**Analytical methods**

Specular X-ray diffraction was carried out with a PANalytical Empyrean diffractometer in θ-θ geometry using CuK$_α$ radiation. On the incident side a parallel beam X-ray mirror was used for monochromatisation and collimation. At the diffracted beam path an anti-scatter slit as well as a 0.02 rad Soller slit was used together with a

PIXcel3D detector operating in either receiving 0D-mode or scanning 1D-mode. Temperature dependent in-situ measurements were performed with a DHS 900 heating stage from Anton Paar Ltd. Graz [21]. The experiments were performed under nitrogen atmosphere. The data are converted into reciprocal space by $q = \frac{4\pi}{\lambda} sin\left(\frac{2\theta}{2}\right) = \frac{2\pi}{d_{hkl}}$ with λ as the wavelength of the primary X-ray beam, $2\theta$ the scattering angle and $d_{hkl}$ the interplanar distance of the (*hkl)* plane. In case of a specular scan the total length of the scattering vector q has a contribution only in z-direction (perpendicular to the substrate surface).

Grazing incidence X-Ray diffraction (GIXD) at ambient temperatures was carried out at the beamline XRD1 at Elettra Synchrotron Trieste with a radiation wavelength of 1.4 Å using an incidence angle of $α_i$ = 0.8° on a goniometer in Kappa geometry [22]. A PILATUS 2M detector was used to collect diffracted intensity. To improve statistics, the sample was rotated during measurement and six images with an exposure time of 30 s each were collected during one full sample rotation. Temperature dependent in-situ grazing incidence X-ray diffraction was performed using a Rigaku SmartLab 9 kW equipped with a Cu rotating anode, collimating parabolic multilayer mirror and pinhole optics. A HyPix 3000 2D detector was utilized for collecting GIXD measurements. Because of low signal a series of 40 images was measured for 1 hour each and summed up to improve the signal to noise ratio. A DHS 1100 commercial high temperature stage from Anton Paar, covered by a carbon dome and filled with a nitrogen atmosphere was used during heating [23].

Data from GIXD are presented as reciprocal space maps with the out-of-plane ($q_z$) and in-plane component ($q_{xy}$) of the scattering vector *q* as an orthogonal basis, with the $q_z$ component parallel to the surface normal and the $q_x / q_y$ plane parallel to the substrate surface. The scattering vector components are determined for each detector pixel from the incident angle α$_i$ and the outgoing angle α$_f$ in the sample coordinate system and from a calibration measurement on an LaB$_6$ film, to determine additional parameters, e.g. the distance from the sample to the detector. Resulting data was evaluated with the use of the in-house developed software package GIDVis [24]. The resulting reciprocal space maps are corrected based on geometrical correction factors, i.e. Lorentz and polarisation factors.

For analysing the films in terms of paracrystallinity, the peaks were fitted with a Gaussian function in both in-plane (q$_{xy}$) and out-of-plane (q$_z$) direction. For the resulting peak widths Δq$_{xy}$ and Δq$_z$ a linear regression based on $(\Delta q)^2 = 4\left[\left(\frac{\pi}{D_{hkl}}\right)^2 + \left(\frac{\pi^3 g^2 m^2}{d_{hkl}}\right)^2\right]$ is calculated, where m and d$_{hkl}$ denote the order of the diffraction peak in terms of its Laue index (*h, k* or *l*) and the inter-planar spacing of the diffraction peaks, respectively. This allows determining the crystallite size D$_{hkl}$ and the amount of paracrystallinity *g* of the sample in two individual directions. No correction for the instrumental peak broadening was used.

**Computational method**

Determination of the molecular packing within the SmE phase was performed by an experimental / computational approach. In the first step the lattice constants were determined by indexing of the GIXD pattern. The crystallographic unit cell was used as an input for Molecular Dynamics (MD) simulation revealing the molecular packing. These simulations were carried out with the LAMMPS software package [25] using the CHARMM general force field version 3.0.1 [26]. Several thousand trial structures are generated, by placing two randomly oriented molecules in an expanded (120%) unit cell. During the simulation run the starting configuration was relaxed and reduced to the experimentally determined unit cell size. The resulting structures are then clustered based on similarities of the packing motifs. Each selected packing motif appeared multiple times at the respective total energy. The obtained packing motifs are judged by a comparison of observed ($F_{obs}$) and calculated structure factors ($F_{calc}$) taken from the diffraction intensities on basis of a reliability factor $R = \frac{\sum ||F_{obs}|-|F_{calc}||}{\sum |F_{obs}|}$.

**Experimental results**

A series of specular X-ray diffraction measurements were performed on a spin coated (5 g/l, 3000 rpm) film in the temperature range from room temperature up to 240°C. Figure 1a shows XRD scans at selected temperatures. At room temperature (25°C) a dominant peak series at $q_z = (0.118 * l)$ Å$^{-1}$ (with $l$ as the Laue index) is observed which can be assigned to the *00l* diffraction peaks of the bulk crystal phase of the molecule Ph-BTBT-10 [12]. The interplanar distance of the *001* plane at 52.3 Å clearly reflects the bilayer structure of the known bulk phase. The absence of *hkl* reflections with $h \neq 0$ or $k \neq 0$ reveals a strong preferred orientation of the crystallites with the (001) plane parallel to the substrates surface. At a temperature of 149°C the peak sequence changes to $q_z = (0.216 * l)$ Å$^{-1}$ which corresponds to the characteristic 00*l* peak series of the SmE phase with an interplanar distance of 29.1 Å of the (001) plane. With increasing temperature up to 215°C the peak position gradually shifts to lower $q_z$ values due to thermal expansion. A sudden shift in the peak positions as well as in the peak intensities is observed at 220°C. This peak series is observed at $q_z = (0.223 * l)$ Å$^{-1}$ corresponding to an interplanar distance of 28.1 Å and is assigned to the *smectic A* (SmA) phase [9][13]. At 240°C, the diffraction peaks disappear which is assigned to a transition to the isotropic state. Figure 1b shows a waterfall plot of the specular diffraction pattern in the temperature range from 130°C to 240°C. Each scan was measured for 10 min at a fixed temperature which was increased in steps of 2°C in the vicinity of the phase transition temperature and in steps of 10°C at more remote temperature ranges. The phase transition temperatures between the bulk phase and the SmE phase is observed at 149°C, the transition to SmA occurs at 220°C and the final melting starts at 235°C. These phase transition temperatures are in quite good agreement with literature [9][10]. The phase transition at 149°C exhibits the vanishing of odd-numbered (*00l*) peaks from the bulk phase and a shift of the leftover even-numbered peaks to smaller q-values. The doubling of peak distances in q-space indicates a reduction of the interplanar distance by a factor of approximately two which means that the bilayer structure changes to a single layer structure [14]. Please note, that the molecular layer thickness corresponds approximately to the length of a single molecule which means that a single molecular sheet is composed by upright standing molecules. Therefore, we can conclude that the SmE phase as well as the SmA phase are formed by single sheets which are stacked upright onto each other parallel to the substrate surface.

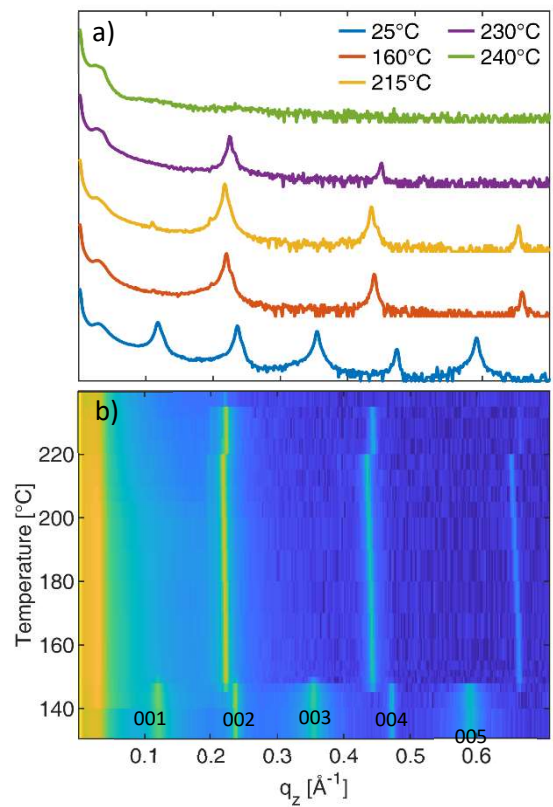

Figure 1. Specular X-ray diffraction of a Ph–BTBT–10 thin film as a function of temperature. Single scans of a film at selected temperatures (a) and a series of scans in the region between 120° and 240°C in a waterfall plot including the three phase transitions: bulk phase / SmE at a temperature of 149°C, SmE / SmA at 220°C and SmA / isotropic state at 235°C (b).

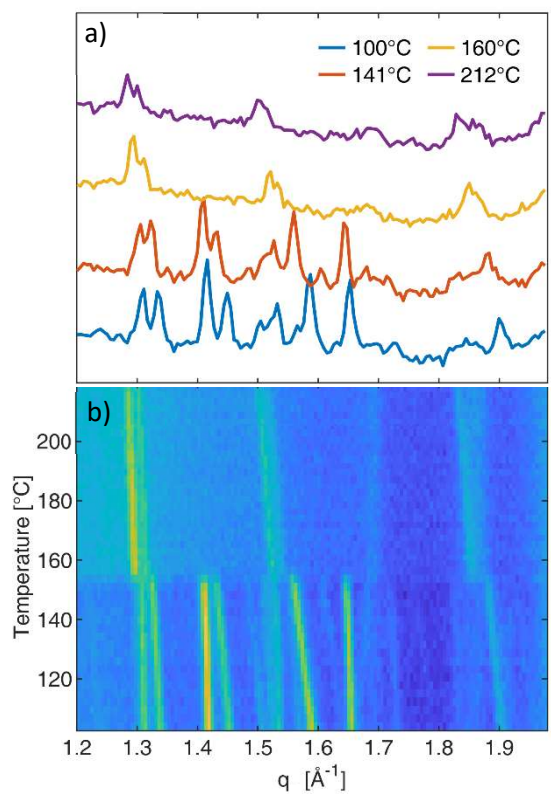

Figure 2. Selected powder diffraction scans at different temperatures (a) and the complete series of X-ray powder diffraction scans in the temperature region between 100°C and 220 °C illustrated in a waterfall plot (b). The phase transition from the bulk phase to the SmE phase is around 153°C.

Figure 2 shows a series of diffraction patterns as a function of temperature measured on a powder sample at higher values of $q$. Selected diffraction patterns are depicted in Figure 2a, while a waterfall plot in a selected temperature range is given in Figure 2b. A clear phase transition is visible at 153°C, a temperature slightly larger than observed at thin films. A continuous shift of the peak pattern is visible also before and after the phase transition due to thermal expansion. To study the evolution of the lattice constants as a function of temperature, the exact peak positions of all observed peaks of the bulk phase (T < 153°C) were determined (namely *00l, 010, 011, 100, 101, 110, 111, 020, 021, 120, 012, 102, 112*). The lattice constants *a, b, c* and the monoclinic angle *ß* were determined as a function of temperature, the result is depicted in Figure 3. Within the bulk structure (temperature range from 60°C to 153°C) the lattice constants *a* and *c* do not change significantly, but the lattice constant *b* increases and the monoclinic angle *ß* decreases.

The diffraction pattern at elevated temperatures (T > 153°C) are characteristic for a SmE phase with dominant peaks at 1.29 Å$^{-1}$, 1.52 Å$^{-1}$ and 1.85 Å$^{-1}$ [27][28]. The refinement of the unit cell parameters after the phase transition resulted in a monoclinic angle very close to 90°, independent of temperature. An orthorhombic unit cell was concluded, a result which is comparable to other SmE phases [29].

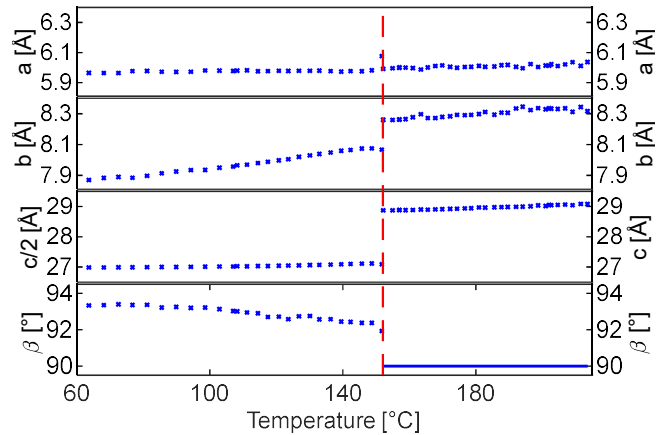

Figure 3. Lattice constants as a function of temperature of the bulk phase below a temperature of 153°C and of the smectic E phase between 153°C and 215°C. In case of the lattice constant c the half of the lattice constant (c/2) is plotted for the bulk phase, while for the SmE phase the full length (c) is plotted.

At the transition temperature of 153°C a clear shift in the lattice constants $b$, $c$ and $β$ are observed, while a roughly continuous behaviour is observed for the lattice constant $a$. The volume of the crystallographic unit cell changes discontinuously from 2488.1 Å$^3$ (corresponding to 622 Å$^3$ per molecule) for the bulk crystal phase to 1462.0 Å$^3$ (731 Å$^3$ per molecule) for the SmE phase.

A more detailed crystallographic study was performed by GIXD. In the first step the bulk structure of Ph-BTBT-10 was investigated for a drop casted film at room temperature. Figure 4a presents an experimental diffraction pattern taken at room temperature, calculated peak positions are given by green spots and peak intensities by the area of the surrounding circles. The characteristic diffraction features of the bulk structure can be clearly identified as enhanced intensities along the *11l*, *02l* and *12l* rod. Moreover, characteristic packing features of the individual bulk structure of Ph-BTBT-10 can be identified by the enhanced intensities of the *112* and *115* peaks.

In a subsequent step the SmE phase was investigated by GIXD at a temperature of 160°C. A spin coated thin film from a 5 g/l solution was used. The corresponding result is shown in Figure 4b-d. A quite different diffraction pattern is observed in comparison to the bulk phase. Only few diffraction peaks are observed, they are arranged in two sequences at different q$_{xy}$ values. Peaks along $q_z = 0$ (in-plane direction) reveal crystallographic order along the substrate surface, these peaks represent crystallographic order within a single smectic plane. The peak series at $q_z \sim 0.22$ Å$^{-1}$ reveals an onset of long range (positional) order perpendicular to the substrate surface (z-direction); these peaks reveal crystallographic order due to stacking of smectic planes. This diffraction pattern is in agreement with previous studies on SmE phases of other molecular materials showing a clear absence of peaks, corresponding to *10l* and *01l* peak series and six dominant peaks between $q = 1.3$ Å$^{-1}$ and 1.9 Å$^{-1}$ (with Laue indices *110, 111, 020, 021, 120, 121*) [30]. Additionally, small intensities are found for the *200, 201* and *112* peaks (compare Fig. 4b-d). The peak indexation is based on the results presented in Figure 3 using the following lattice constants of the orthorhombic unit cell at a temperature of 160°C: *a = 6.051 Å, b = 8.303 Å* and *c = 29.10 Å*.

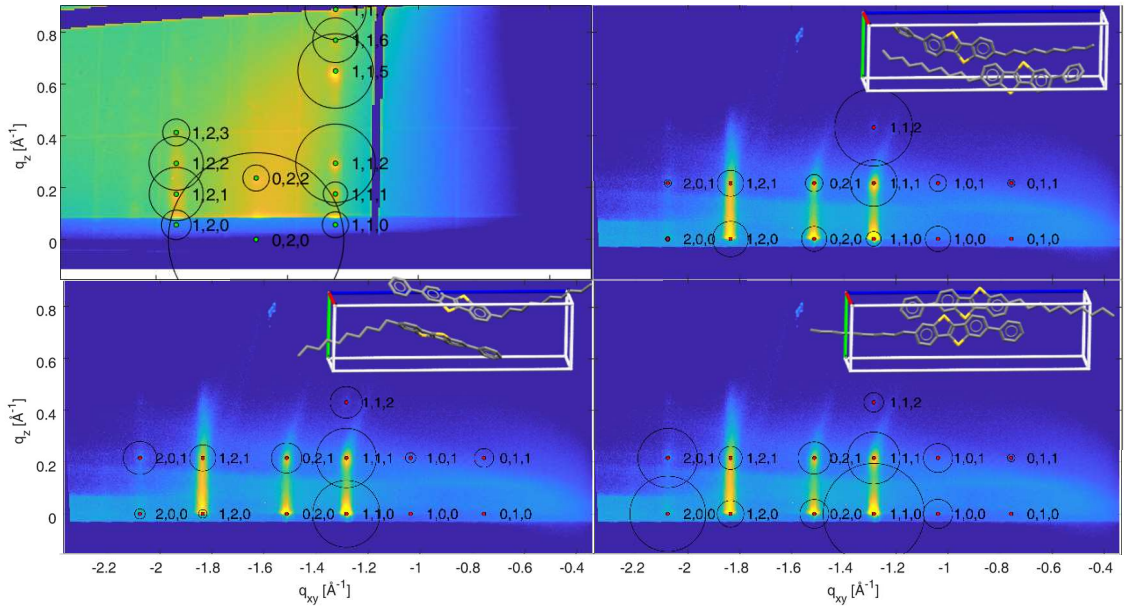

Figure 4. Grazing incidence X-ray diffraction pattern of Ph-BTBT-10 thin films at room temperature (a) and at a temperature of 160°C (b-d) with logarithmic scaling of the intensity. Additionally, calculated diffraction patterns are shown with points and circles representing positions and intensities of the calculated Bragg peaks, respectively. Different packing motifs are selected: the crystallographic bulk phase (a), Sm*E* with mixed layers (b), Sm*E* with nano-segregated layers possessing a herringbone arrangement (c) and a stacked arrangement (d). The insets give the different packing motifs relative to the crystallographic unit cell.

Since only a limited number of peaks are observed, the long range order of the molecules within the SmE phase is far from perfect. But in several cases higher order diffraction peaks are observed, so that an analysis in terms of a paracrystalline model can be applied [31]. Figure 5 shows fits of the paracrystalline model for five families of diffraction peaks (*11l, 02l* and *20l, h20, 1k0*) in terms of their in-plane peak widths $\Delta q_{xy}$ and out-of-plane peak widths $\Delta q_z$, respectively, with a summary of the fit parameters in Table 1. Although the number of data points are limited (two or three), the complete set of data gives a clear trend for the two different directions. A crystallite size of about $D_z = 100$ Å and a paracrystallinity $g_z = 10\%$ is found for the out-of plane direction while values $D_{xy} = 50$ Å and $g_{xy} = 4\%$ is found for the in-plane direction. A g-value of smaller than 10% is classified as diffuse scattering from a disordered state [32].

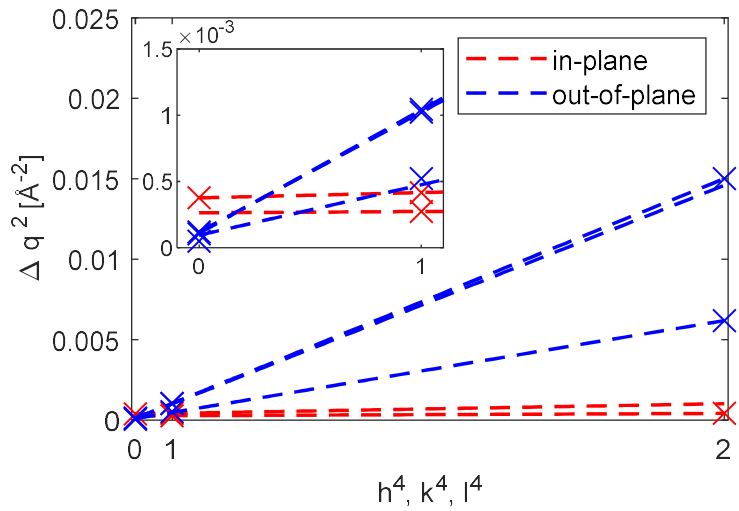

Figure 5. The square of in-plane peak width ($\Delta q_{xy}^2$) and of the out-of-plane peak width ($\Delta q_z^2$) plotted as a function of the fourth order of the Laue indices $h,k,l$. The linear regression is based on the paracrystalline model. The inset shows the range between $h,k,l = 0$ and $h,k,l = 1$.

| Laue indices | $D_z$ [Å] | $D_{xy}$ [Å] | $g_z$ [%] | $g_{xy}$ [%] |
|---|---|---|---|---|
| 1 1 l | 96.7 | - | 12.3 | - |
| 0 2 l | 103 | - | 9.01 | - |
| 2 0 l | 91.8 | - | 11.2 | - |
| h 2 0 | - | 51.5 | - | 5.6 |
| 1 k 0 | - | 61.5 | - | 3.5 |

Table 1: Crystallite sizes $D_z$ and $D_{xy}$ and degrees of paracrystallinity $g_x$ and $g_{xy}$ for five sets of diffraction peaks determined on basis of the paracrystalline model for the *smectic E* phase of Ph-BTBT-10. The respective values for the in-plane direction (along the smectic layers) and for the out-of-plane direction (across the smectic layers) are denoted by the superscripts *z* and *xy*, respectively.

In a next step, the molecular packing within the SmE phase is analysed. The sudden change of the lattice constant *c* at the phase transition temperature from 54.2 Å for the bulk phase to 28.9 Å in the SmE phase (compare Fig.3) reflects the transition from a bilayer structure with *head-to-head* arrangement to a single layer structure with *head-to-tail* arrangement. The transition can happen by lateral and vertical displacements of the molecule that avoid an energy demanding complete flip of the whole molecule [14][16]. As a consequence, the single sheet structure is formed by anti-parallel molecules with an alternating orientation of the aromatic cores giving the alkyl chains a parallel and antiparallel alignment to the c*-axis (perpendicular to the smectic layers).

The molecular dynamics simulations are based on the crystallographic unit cell containing two molecules. A number of different candidate packing motifs are found. Three of them show anti-parallel molecules, they were selected for further considerations. Their respective molecular packings are shown as insets of Figure 4b-d. In case of the first packing motif (denoted as mixed layers, Fig. 4b), the individual layers are composed by blended decyl chains and aromatic units. In case of the second and third packing motif (denoted as nano-segregated), the structure shows a separation of the molecular parts into layers of aromatic units and layers of decyl chains. Within the nano-segregation further differentiation is found for the packing of the aromatic units i) tilting relative to each other (herringbone arrangement, Fig. 4c) or ii) parallel stacking upon each other (stacked arrangement, Fig. 4d). The respective crystallographic information files (cif - files) of these three packing motifs are given in the Supporting Information.

In a subsequent step, the three different packing motifs were used to calculate structure factors and compare them with experimental values. In the first step R-factors were calculated for all observed Bragg peaks (compare Fig.4b-d), the results are shown in Table 2. The largest R-factor is found for the mixed layer system, both nano-segregated structures show considerably smaller R-values. From this point of view, the mixed layer motif seems to be the least matching molecular packing within the SmE phase. However, it cannot be concluded which type of nano-segregated structure is present.

| packing motif | R | Energy |
|---|---|---|
| mixed layer | 1.53 | 130.7% |
| nano-segregated herringbone | 0.67 | 100% |
| nano-segregated stacking | 0.65 | 159.5% |

Table 2: Comparison of the reliability factors for X-ray diffraction (R-factors) and of the total energies obtained from Molecular Dynamics calculations for the three different packing motifs. The energies are given relative to the lowest obtained energy.

Our molecular dynamics simulations reveal that the three different packing motifs are associated with characteristic energies. Therefore, we use these energies as a criterium for assigning a specific packing motif to the SmE phase. Since the obtained total energies are less meaningful, we decided to give fractional values relative to the lowest energy state (Tab.2). Our calculation reveals that the herringbone packing is the lowest energy state and the difference with respect to the other two packings is comparably large. Please note that the higher energies for the two other packing motifs (mixed layer / nano-segregation with parallel stacking) is a consequence of squeezing the molecules into a given crystallographic unit cell. Therefore, the given energy values are not representative for a relaxed state of these two packing motifs.

Based on the comparison of the experimental and the calculated diffraction pattern together with energy considerations of the Molecular Dynamics simulations, we conclude that nano-segregation with herringbone packed aromatic units is present within the SmE phase of the molecule Ph-BTBT-10. The result that herringbone packing is present within the SmE phase is in accordance with the literature [26][28][34]. Please note that the negligible intensities of the *10l* and *011* peaks in the experimental diffraction pattern (compare Fig. 5c) is a characteristic feature of herringbone packing, frequently observed even for alkyl terminated conjugated molecules [17][33].

The analysis of the molecular packing for the nano-segregated herringbone structure in terms of molecular conformation reveals that a planar conformation of the aromatic unit is obtained. This can be expected, since the anti-parallel alignment of neighbouring the molecules causes that the phenyl and the BTBT units are next to each other. Thus, an integration of both units into an intermolecular herringbone arrangement makes a planarization reasonable. A herringbone angle of 54° is obtained which is comparable with an angle of 50° observed for the bulk phase. The tilt angle of the decyl side chain relative to the aromatic unit is 40°, this value is comparable to the 35° tilt angle observed within the bulk phase.

**Discussion**

Asymmetric molecules which are composed of aromatic units and aliphatic chains show often combined SmE and SmA phases [29][27]. In the case of the molecule Ph-BTBT-10 we found a stacking distance of 28.1 Å for the SmA phase which is in quite good agreement with atomistic modelling by molecular dynamics [15]. Previous analysis of the molecular packing within the SmE phase of the molecule Ph-BTBT-10 was based on the transition from the bulk phase to the SmE phase. A bilayer structure showing *head-to-head* arrangements of the molecules is transferred to the SmE phase with a single layer structure with *head-to-tail* arrangement. A mechanism for a defined transition from a *head-to-head* arrangement to a *head-to-tail* arrangement is possible by collective translational movement of the molecules [14][16].

Our approach with molecular dynamics simulation is based on the experimentally observed crystallographic unit cell of the SmE phase. Starting from random arrangements of the molecules, the system is reduced to the actual unit cell size. The obtained packing motifs are analysed in terms of energy and in accordance with the experimental diffraction pattern. Please note that the presented results consider rigid aromatic units and alkyl chains. Thermal movements, molecular disorder and displacement of the molecules (and atoms) is not included. It is stated in the literature that the alkyl chains are in a molten state within the SmE phase and the molecular packing is determined by the rigid character of the aromatic units[34][35]. We found that the packing of the aromatic units appears in herringbone arrangement.

Our results reveal that the molecular packing of the molecules is rather nano-segregated which means that the phenyl-BTBT units and the decyl units form separated layers within a molecular sheet. This contradicts the classical view of a SmE phase, but it is in agreement with more recent concepts [36][37]. The nano-segregated structure of the SmE phase together with the anti-parallel alignment of the molecules result in an interdigidation of the decyl chains originating from two aromatic layers of neighbouring layers. Interdigidation of alkyl chains within sheet structures may be favoured from the thermodynamic state, but may also be hindered by growth kinetics as discussed in detail for the polymer poly(3-hexylthiophene) [38][39]. It is found also for another BTBT based molecule, e.g. 2,7-dioctyloxy[1]benzothieno[3,2-b]benzothiophene that polymorph phases with and without interdigidated side chains exists [40].

## Conclusion

Thin films of the molecule Ph-BTBT-10 are investigated in terms of their molecular packing as a function of temperature. A phase transition from the bulk state to a SmE phase and finally to a SmA phase is observed by X-ray diffraction experiments. Sharp transitions from the bulk crystal phase to the SmE phase and further to SmA phase are observed at 149°C and 220°C, respectively. The transition to the SmE phase is associated with a change from a monoclinic to an orthorhombic unit cell with a discontinuous change of the lattice constants. The available space of a single molecule increases substantially from 622 Å$^3$ in the bulk phase to 731 Å$^3$ in the SmE phase. Analysis of the peak width by a paracrystalline model reveals that the in-plane order is reduced (crystal size of 50 Å and 4% paracrystallinity) in comparison to the out-of plane order (100 Å / 10%). Based on the crystallographic lattice constants, obtained from GIXD analysis, molecular packing analysis was performed by molecular dynamics calculations. Three different packing motifs with antiparallel alignment of the molecules are found. A comparison of calculated and experimental diffraction patterns reveals that the "mixed layer" packing motif is unlikely while the nano-segregation of decyl chains from the aromatic units is more probably present within the SmE phase. Considering the two different types of aromatic packing within nano-segregation ("parallel stacking" and "herringbone arrangement"), the lowest energy in the molecular dynamics simulations is found for herringbone arrangement of the aromatic units.

## Associated Content

Crystallographic information files for the three considered different molecular packings. Structures formed by i) sheets of blended decyl chains and aromatic units (mixed_layer.cif), sheets of nano-segregated decyl chains and aromatic units with ii) herringbone arrangement of the aromatic units (nanoegregated_herringbone.cif) and iii) with parallel stacking of the aromatic units (nano-segregated_stacking.cif).

## Acknowledgments

The authors acknowledge the synchrotron Elettra, Trieste for allocation of synchrotron radiation and thank Luisa Barba and Nicola Demitri for assistance in using beamline XRD1. The measurements at CzechNanoLab Research Infrastructure were financially supported by MEYS CR (LM2018110).

## Disclosure statement

No potential conflict of interest was reported by the authors.

## Funding

The work was supported by the Austrian Science Foundation (FWF) under Grant [P30222]; CzechNanoLab Research Infrastructure under Grant [MEYS CR-LM2018110].

# References

[1] Ebata H, Izawa T, Miyazaki E, et al. Highly soluble [1]benzothieno[3,2-b]benzothiophene (BTBT) derivatives for high-performance, solution-processed organic field-effect transistors. J. Am. Chem. Soc. 2007;129:15732–15733.

[2] Minemawari H, Yamada T, Matsui H, et al. Inkjet printing of single-crystal films. Nature. 2011;475:364–367.

[3] Izawa T, Miyazaki E, Takimiya K. Molecular ordering of high-performance soluble molecular semiconductors and re-evaluation of their field-effect transistor characteristics. Adv. Mater. 2008;20:3388–3392.

[4] Dohr M, Werzer O, Shen Q, et al. Dynamics of monolayer-island transitions in 2,7-dioctyl- benzothienobenzthiophene thin films. ChemPhysChem. 2013;14:2554–2559.

[5] Spreitzer H, Kaufmann B, Ruzié C, et al. Alkyl chain assisted thin film growth of 2,7-dioctyloxy-benzothienobenzthiophene. J. Mater. Chem. C. 2019;7:8477–8484.

[6] Lyu L, Niu D, Xie H, et al. The correlations of the electronic structure and film growth of 2,7-diocty[1]benzothieno[3,2-: B] benzothiophene (C8-BTBT) on SiO2. Phys. Chem. Chem. Phys. 2017;19:1669–1676.

[7] Iino H, Hanna JI. Availability of liquid crystallinity in solution processing for polycrystalline thin films. Adv. Mater. 2011;23:1748–1751.

[8] Iino H, Hanna JI. Liquid crystalline organic semiconductors for organic transistor applications. Polym. J. 2017;49:23–30.

[9] Iino H, Usui T, Hanna JI. Liquid crystals for organic thin-film transistors. Nat. Commun. 2015;6:6828.

[10] Inoue S, Minemawari H, Tsutsumi J, et al. Effects of Substituted Alkyl Chain Length on Solution-Processable Layered Organic Semiconductor Crystals. Chem. Mater. 2015;27:3809–3812.

[11] Wu H, Iino H, Hanna JI. Thermally induced bilayered crystals in a solution-processed polycrystalline thin film of phenylterthiophene-based monoalkyl smectic liquid crystals and their effect on FET mobility. RSC Adv. 2017;7:56586–56593.

[12] Minemawari H, Tsutsumi J, Inoue S, et al. Crystal structure of asymmetric organic semiconductor 7-decyl-2-phenyl[1]benzothieno[3,2-b][1]benzothiophene. Appl. Phys. Express. 2014;7:2–5.

[13] Iino H, Hanna JI. Liquid crystal and crystal structures of a phenyl-benzothienobenzothiophene derivative. Mol. Cryst. Liq. Cryst. 2017;647:37–43.

[14] Yoneya M. Monolayer Crystal Structure of the Organic Semiconductor 7-Decyl-2-phenyl[1]benzothieno[3,2- b][1]benzothiophene. J. Phys. Chem. C. 2018;122:22225–22231.

[15] Baggioli A, Casalegno M, Raos G, et al. Atomistic Simulation of Phase Transitions and Charge Mobility for the Organic Semiconductor Ph-BTBT-C10. Chem. Mater. 2019;31:7092–7103.

[16] Yoneya M. Monolayer crystal structure of the organic semiconductor 7-decyl-2-phenyl[1]benzothieno[3,2-b][1]benzothiophene, revisited. Jpn. J. Appl. Phys. 2020;59.

[17] Lercher C, Röthel C, Roscioni OM, et al. Polymorphism of dioctyl-terthiophene within thin films: The role of the first monolayer. Chem. Phys. Lett. 2015;630:12–17.

[18] Pithan L, Nabok D, Cocchi C, et al. Molecular structure of the substrate-induced thin-film phase of tetracene. J. Chem. Phys. 2018;149.

[19] Jones AOF, Röthel C, Lassnig R, et al. Solution of an elusive pigment crystal structure from a thin film: a combined X-ray diffraction and computational study. CrystEngComm. 2017;19:1902–1911.

[20] Sanzone A, Mattiello S, Garavaglia GM, et al. Efficient synthesis of organic semiconductors by Suzuki-Miyaura coupling in an aromatic micellar medium.

Green Chem. 2019;21:4400–4405.

[21] Resel R, Tamas E, Sonderegger B, et al. A heating stage up to 1173 K for X-ray diffraction studies in the whole orientation space. J. Appl. Crystallogr. 2003;36:80–85.

[22] Fumagalli E, Campione M, Raimondo L, et al. Grazing-incidence X-ray diffraction study of rubrene epitaxial thin films. J. Synchrotron Radiat. 2012;19:682–687.

[23] DHS1100: A new high-temperature attachment for materials science in the whole orientation space. J. Appl. Crystallogr. 2007;40:202.

[24] Schrode B, Pachmajer S, Dohr M, et al. GIDVis: a comprehensive software tool for geometry-independent grazing-incidence X-ray diffraction data analysis and pole-figure calculations. J. Appl. Crystallogr. 2019;52:683–689.

[25] Plimpton S. Fast parallel algorithms for short-range molecular dynamics. J. Comput. Phys. [Internet]. 1995;117:1–19. Available from: http://www.cs.sandia.gov/~sjplimp/main.html.

[26] Brooks BR, Brooks C., Mackerell AD, et al. CHARMM: Molecular dynamics simulation package. J. Comput. Chem. 2009;30:1545–1614.

[27] Pensec S, Tournilhac FG, Bassoul P. Structure of the mesophases formed by a perfluoroalkyl/biphenyl compound. Amphiphilic and steric effects. J. Phys. II. 1996;6:1597–1605.

[28] Jasiurkowska M, Budziak A, Czub J, et al. X-ray studies on the crystalline e phase of the 4-n-alkyl-4′- isothiocyanatobiphenyl homologous series (nBT, n = 2-10). Liq. Cryst. 2008;35:513–518.

[29] Diele S, Tosch S, Mahnke S, et al. Structure and Packing in Smectic E and Smectic A Phases in the Series of 4-n-Alkyloxy-4′-alkanoylbiphenyls. Cryst. Res. Technol. 1991;26:809–817.

[30] Seddon JM. Structural Studies of Liquid Crystals by X-Ray Diffraction. Handb. Liq. Cryst. Set. 2008;635–679.

[31] Hosemann R. Crystalline and paracrystalline order in high polymers. J. Appl. Phys. 1963;34:25–41.

[32] Bonart R, Hosemann R, McCullough RL. The influence of particle size and distortions upon the X-ray diffraction patterns of polymers. Polymer (Guildf). [Internet]. 1963;4:199–211. Available from: http://www.sciencedirect.com/science/article/pii/0032386163900260.

[33] Jones AOF, Chattopadhyay B, Geerts YH, et al. Substrate-induced and thin-film phases: Polymorphism of organic materials on surfaces. Adv. Funct. Mater. 2016;26:2233–2255.

[34] Yamamura Y, Adachi T, Miyazawa T, et al. Calorimetric and spectroscopic evidence of chain-melting in smectic e and smectic a phases of 4-alkyl-4′-isothiocyanatobiphenyl (n TCB). J. Phys. Chem. B. 2012;116:9255–9260.

[35] Horiuchi K, Yamamura Y, Pełka R, et al. Entropic contribution of flexible terminals to mesophase formation revealed by thermodynamic analysis of 4-Alkyl-4′-isothiocyanatobiphenyl (n TCB). J. Phys. Chem. B. 2010;114:4870–4875.

[36] Saito K, Miyazawa T, Fujiwara A, et al. Reassessment of structure of smectic phases: Nano-segregation in smectic E phase in 4-n-alkyl-4-isothiocyanato-1,1-biphenyls. J. Chem. Phys. 2013;139.

[37] Miyazawa T, Yamamura Y, Hishida M, et al. Revisiting smectic E structure through swollen smectic E phase in binary system of 4-nonyl-4′-isothiocyanatobiphenyl (9TCB) and n-nonane. J. Phys. Chem. B. 2013;117:8293–8299.

[38] Himmelberger S, Duong DT, Northrup JE, et al. Role of side-shain branching on thin-film structure and electronic properties of polythiophenes. Adv. Funct. Mater. 2015;25:2616–2624.

[39] Casalegno M, Nicolini T, Famulari A, et al. Atomistic modelling of entropy driven phase transitions between different crystal modifications in polymers: The case of poly(3-alkylthiophenes). Phys. Chem. Chem. Phys. 2018;20:28984–

28989.

[40]  Schrode B, Jones AOF, Resel R, et al. Substrate-Induced Phase of a Benzothiophene Derivative Detected by Mid-Infrared and Lattice Phonon Raman Spectroscopy. ChemPhysChem. 2018;19:993–1000.

TOC (table of content figure)

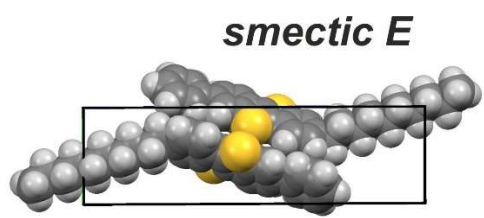